\documentclass[aps,pre,twocolumn,superscriptaddress]{revtex4}
\usepackage{epsfig}
\usepackage{times}
\usepackage{color}
\usepackage[dvipdfm]{hyperref}
\hypersetup{colorlinks=false,pdfborder=0}

\begin{document}

\title{Understanding recurrent crime as system-immanent collective behavior}

\author{Matja{\v z} Perc}
\thanks{Electronic address: \href{mailto:matjaz.perc@uni-mb.si}{\textcolor{blue}{matjaz.perc@uni-mb.si}};
URL: \href{http://www.matjazperc.com/}{\textcolor{blue}{www.matjazperc.com}}}
\affiliation{Faculty of Natural Sciences and Mathematics, University of Maribor, Koro{\v s}ka cesta 160, SI-2000 Maribor, Slovenia}

\author{Karsten Donnay}
\affiliation{ETH Zurich, CLU E1, Chair of Sociology, in particular of Modeling and Simulation, Clausiusstrasse 50, CH-8092 Zurich, Switzerland}

\author{Dirk Helbing}
\affiliation{ETH Zurich, CLU E1, Chair of Sociology, in particular of Modeling and Simulation, Clausiusstrasse 50, CH-8092 Zurich, Switzerland}

\affiliation{Risk Center, ETH Zurich, Swiss Federal Institute of Technology, Scheuchzerstrasse 7, CH-8092 Zurich, Switzerland}

\begin{abstract}
Containing the spreading of crime is a major challenge for society. Yet, since thousands of years, no effective strategy has been found to overcome crime. To the contrary, empirical evidence shows that crime is recurrent, a fact that is not captured well by rational choice theories of crime. According to these, strong enough punishment should prevent crime from happening. To gain a better understanding of the relationship between crime and punishment, we consider that the latter requires prior discovery of illicit behavior and study a spatial version of the inspection game. Simulations reveal the spontaneous emergence of cyclic dominance between ``criminals'', ``inspectors'', and ``ordinary people'' as a consequence of spatial interactions. Such cycles dominate the evolutionary process, in particular when the temptation to commit crime or the cost of inspection are low or moderate. Yet, there are also critical parameter values beyond which cycles cease to exist and the population is dominated either by a stable mixture of criminals and inspectors or one of these two strategies alone. Both continuous and discontinuous phase transitions to different final states are possible, indicating that successful strategies to contain crime can be very much counter-intuitive and complex. Our results demonstrate that spatial interactions are crucial for the evolutionary outcome of the inspection game, and they also reveal why criminal behavior is likely to be recurrent rather than evolving towards an equilibrium with monotonous parameter dependencies.
\end{abstract}

\maketitle

\section{Introduction}

Crime may be seen as human activity that deviates from social norms in intolerable ways. Despite our best efforts to fight it, crime continues to plague society since thousands of years and endanger its very foundation: social order. To contain criminal activity, societies enforce social norms by ``pool punishment'' (through institutions such as the police) \cite{gurerk_s06, sigmund_n10, traulsen_prsb12} or ``peer punishment'' (i.e. decentralized, individual sanctioning efforts) \cite{fehr_aer00, gardner_a_an04, rockenbach_n06, eldakar_pnas08, boyd_s10}. In this paper, we will focus on peer punishment, which has also been discussed as ``altruistic punishment'' \cite{fehr_n02, fowler_n05, fowler_pnas05, lehmann_an07} or ``costly punishment'' \cite{fehr_n03, boyd_pnas03, henrich_s06b, milinski_n08}. In contrast to previous work, however, we will take into account that the detection of crime requires a costly inspection effort.

\begin{figure}
\centerline{\epsfig{file=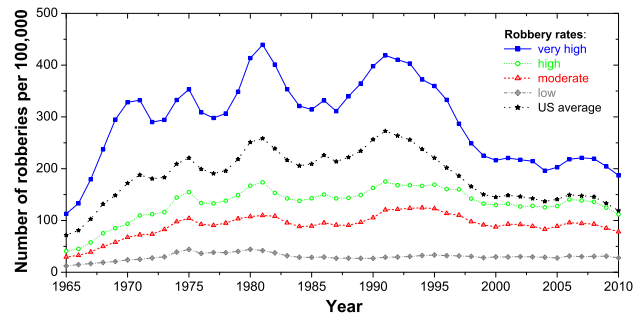,width=8.3cm}}
\caption{Empirical evidence for the recurrent nature of crime, depicting the number of robberies per 100,000 population. Next to the average for the United States (black), the figure shows averages of states with very high (blue), high (green), moderate (red) and low (grey) robbery rates. The four categories are obtained through grouping states by their average robbery rate over the period 1965-2010 and then assigning the 25\% most affected states to the first category, the 25\% next most affected states to the second category, etc. Examples of states with very high robbery rates are Washington D.C. and New York. States with low robbery rates are, for instance, North Dakota and Utah. In states with very high robbery rates, the number of robberies oscillates considerably over time. In states with high and moderate robbery rates numbers follow the same up and down trends but at significantly lower levels and with smaller variations. In the states with low robbery rates numbers are very low and remain more or less stable over time. Similar dependencies are found for other kinds of crime such as motor vehicle theft and property crime.}
\label{empirical}
\end{figure}

Empirical data show that crimes, regardless of type and severity, are often recurrent \cite{UCR,handbook}. The robbery rate in the United States, for example, began to rise steadily in the 1960s and oscillated at around 200 robberies per 100,000 for 20 years. It only started to significantly decline in the 1990s (see Fig.~\ref{empirical}). Notably, the significantly reduced robbery rates in the United States after the year 2000 remain higher than in many other Western democracies but are much lower than in Belgium or Spain, the countries experiencing the highest robbery rates worldwide with well more than 1000 robberies per 100,000 \cite{UNODC2}.

Although large efforts are invested to deter and prevent crime, it reappears over and over
again, sometimes in more vigorous forms than before. According to rational
choice theories of crime, strong enough punishment should be able to prevent it. Yet
for thousands of years, and regardless of how strong sanctions are,
they fail to prevent crime. Until today, a fully satisfactory explanation of the
mechanism behind the recurrent nature of crime is still lacking.

In fact, researchers have struggled to find consistent explanations for the recurrence of crime and its trends in the past 50 years. Beside structural factors, such as unemployment \cite{Raphael_2001,Lin_2008} and economic deprivation \cite{LaFree_1999}, studies have highlighted the influence of demography \cite{Zimring_2007}, youth culture \cite{Curtis_1998, Johnson_2006}, social institutions \cite{LaFree_1998} and urban development \cite{Barker_2012}, but also of political legitimacy \cite{LaFree_1999},  law enforcement strategies \cite{Corsaro_2009, McGarrell_2010}, and the criminal justice system \cite{Zimring_2007}. Recent work in criminology emphasizes that trends in the levels of crime may be best understood as arising from a complex interplay of these factors \cite{Barker_2012, Gomez_2006}.

In this paper, we adopt this point of view and show that fighting crime is in fact not a simple gain-loss type of activity which can be understood by ``linear thinking''. Increasing one factor will not necessarily lead to a monotonous change in another, but may have a rather different, counter-intuitive or hardly foreseeable impact. Furthermore, note that crime patterns feature typical characteristics of complex system behavior. Commentators agree that the same general set of factors affected crime rates in the past 50 years, yet the crime statistics show significantly different phases throughout that period. Changes from one phase to the next occur suddenly: for example, nobody expected the decline in the 1990s. In fact, analysts predicted rising crime rates \cite{Wilson_1995, Fox_1996}.

Recognizing the complex interactions of crime and its sanctioning, is it possible to understand why increased punishment fines do not necessarily reduce crime rates \cite{doob_cj03}, as one might expect \cite{becker_jpe68}? For example, why do murders still happen in countries with death penalty, despite the high discovery rate of murderers?

The occurrence of anti-social punishment in certain countries adds further empirical evidence to the counter-intuitive nature of punishment efforts \cite{herrmann_s08}. For these and further reasons, some scientists even question the effectiveness of punishment strategies in general and have suggested that rewards for desirable behaviors might in fact be more effective \cite{dreber_n08, rand_s09, rand_nc11}.

To some extend, crime might be considered as a social dilemma situation. It would be favorable for all, if nobody committed a crime, but there are individual incentives to do so. As large-scale corruption and tax evasion in some troubled countries or mafia and drug wars show, ``tragedy of the commons'' are actually possible \cite{hardin_g_s68}. In fact, if nobody is watching, criminal behavior (such as stealing) may seem rational, since it promises ample reward $g>0$ for little effort. To change this, it appears logical to modify the decision situation through a fine $f>0$ in such a way that the probability $p$ of catching a criminal times the imposed fine $f$ eliminates the reward of the crime, i.e.
\begin{equation}
g - pf < 0 \, .
\label{one}
\end{equation}
However, the success of this strategy depends on the probability of catching a criminal, and this requires an inspection effort. With this in mind, we here study criminal activity in the context of a simple, evolutionary game-theoretical model \cite{short_pre10} based on the inspection game---a well established model in the sociological literature for the dynamics of crime \cite{tsebelis_rs90, inspect}. The game addresses the question of why anybody would be willing to invest into costly inspection activity. The problem is related to a ``second-order free-rider dilemma'' \cite{panchanathan_n04, fowler_n05b}, as individuals are tempted to benefit from the inspection activities of others without contributing to them.

In order to overcome the second-order free-rider dilemma \cite{helbing_ploscb10}, our contribution analyzes the effect of spatial interactions \cite{nowak_n92b, nowak_pnas94, nowak_ijbc94, hauert_n04, santos_pnas06, ohtsuki_n06, szabo_pr07, tanimoto_pre07, santos_n08, fu_pre09, tarnita_jtb09, tanimoto_pre12} in the inspection game. Introducing this aspect has a profound impact on the overall dynamics. The original version of the inspection game implies that stronger punishment would decrease inspection rates rather than crime. Compared to this, studying the spatiotemporal dynamics of crime \cite{short_pnas10, short_siam10, brantingham_cr12}, as we do it, gives a somewhat different and more differentiated picture. Taking into account spatial interactions, i.e. the fact that not everybody is connected with everybody else, reveals the origin of the recurrent nature of crime. Our results demonstrate that crime is indeed expected to be recurrent as long as there is a gain associated with criminal activity. Furthermore, by systematically exploring the parameter dependencies, we find different kinds of possible outcomes (``phases''), some of which display a very interesting dynamics. In particular, we reveal continuous and discontinuous phase transitions to different final states.

The level of complexity governing criminal activities in competition with sanctioning efforts appears to be much greater than it has been assumed so far. Our results may help to explain the counterintuitive impact of punishment on the occurrence of crime. To interpret our results, it is important to consider intricate spatial interaction patterns. The emergent collective behaviors cannot be predicted solely by looking at the interactions between individual agents, but must be attributed to forces of self-organization, which suddenly appear when critical parameter thresholds are crossed.

\section{Results}

\subsection{Spatial inspection game}
We study the spatial inspection game on a fully occupied $L\times L$ square lattice with periodic boundary conditions. Simulated individuals (so-called agents) play the game with their $k=4$ nearest neighbors. Alternatively, we consider regular small-world graphs where a fraction $\lambda$ of all links is randomly rewired. The game involves three kinds of strategies $s_x$, between which the players $x$ are assumed to change, depending of the success they expect from their respective strategy. These behavioral strategies are those of ``criminals'' ($s_x = C$), punishing ``inspectors'' ($s_x = P$), and ``ordinary individuals'' ($s_x =O$), who neither commit crimes nor participate in inspection activities.

Ordinary people receive no payoffs when encountering inspectors or other ordinary individuals. Only when faced with criminals, they suffer the consequences of crime in form of a negative payoff $-g \le0$. The criminals, when facing ordinary individuals, make the equivalent gain $g\ge 0$. When facing inspectors, however, criminals obtain the payoff $g-f$, where $f\ge 0$ is a punishment fine. If faced with each other, none of two interacting criminals is assumed to have a benefit. Inspectors, on the other hand, always have the cost of inspection, $c \ge 0$, but when confronted with a criminal, an inspector receives the reward $r\ge 0$, i.e. the related payoff is $r-c$.

According to the standard parametrization of the inspection game, we have four parameters that determine the set-up. For convenience, however, and in order to decrease the dimensionality of the parameters space, we re-scale the parameters to obtain three dimensionless parameters characterizing the inspection game. These are the ``(relative) inspection costs'' $\alpha=c/f$, which determine how costly inspection is compared to the imposed fines, the ``(relative) temptation'' $\beta=g/f$, which determines how tempted individuals are to commit crimes, and the ``(relative) inspection incentive'' $\gamma=r/f$, which determines the efficiency of sanctioning efforts. If $\gamma<1$, the fines imposed on criminals are not completely turned into rewards, while $\gamma>1$ means that inspection is subsidized by the state or creates some interest rates or synergy effects.
In terms of the parameters $\alpha$, $\beta$ and $\gamma$, the scaled payoffs of ordinary individuals, inspectors and criminals can be expressed as:
\begin{eqnarray}
\pi_O &=& - \beta N_C , \\
\pi_P &=& \gamma N_C -\alpha, \\
\pi_C &=& \beta N_O + (\beta - 1) N_P .
\end{eqnarray}
Here $N_O$, $N_P$ and $N_C$ are the numbers of ordinary individuals, inspectors and criminals among the $k=4$ nearest neighbors.

\subsection{Evolutionary dynamics}

\begin{figure}
\centerline{\epsfig{file=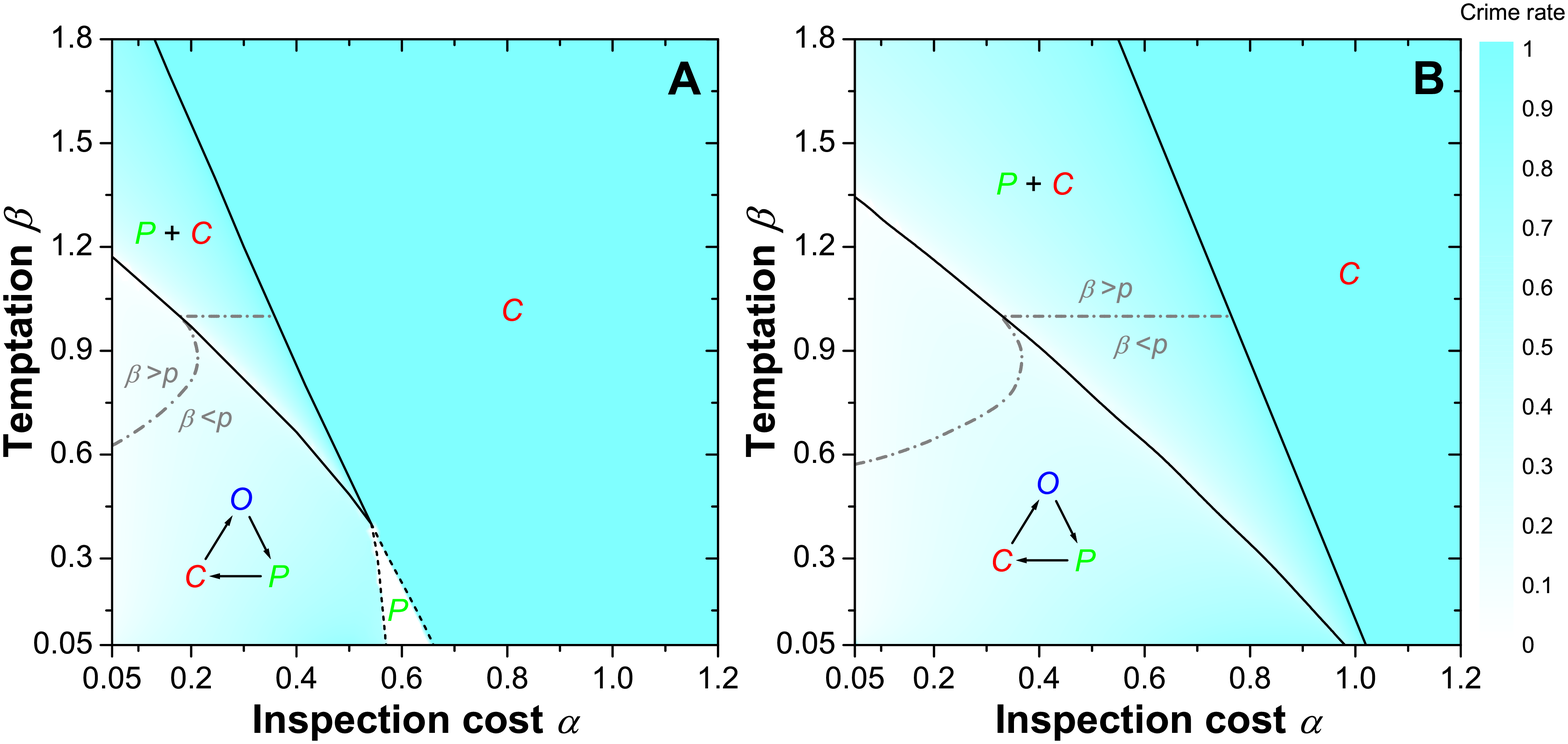,width=9cm}}
\caption{Phase diagrams, demonstrating the spontaneous emergence and stability of the recurrent nature of crime and other possible outcomes of the evolutionary competition of criminals ($C$), ordinary individuals ($O$) and punishing inspectors ($P$). The diagrams show the strategies remaining on the square lattice after sufficiently long relaxation times as a function of the (relative) inspection costs $\alpha$ and the (relative) temptation $\beta$, (\textbf{A}) for inspection incentive $\gamma=0.5$ and (\textbf{B}) for $\gamma=1.0$. The overlayed color map encodes the crime rate, i.e. the stationary density of criminals in the system. (\textbf{A}) For small and intermediate values of $\alpha$ and $\beta$, cyclic dominance between the three strategies characterizes the evolutionary dynamics. Criminals outperform ordinary people, ordinary people outperform inspectors, and inspectors outperform criminals. This cyclic dominance leads to recurrent outbreaks of crime during the evolutionary process. If either $\alpha$ or $\beta$ exceed a certain threshold, the cyclic phase ends with a continuous phase transition to a mixed $P+C$ phase (lower solid line), where inspectors and criminals coexist. Further increasing the two parameters leads to another continuous transition (upper solid line) and an absorbing $C$ phase, where criminals dominate. In other words, when a certain value of temptation $\beta$ is exceeded, it cannot be compensated by larger inspection costs $\alpha$ anymore. A re-entry into the cyclic $C+O+P$ phase is possible through a succession of two discontinuous phase transitions (dashed lines) occurring for sufficiently small $\beta$ and decreasing inspection costs. First, the absorbing $C$ phase changes abruptly to an absorbing $P$ phase dominated by inspectors, which then changes abruptly to the cyclic phase. (\textbf{B}) Increasing the value of $\gamma$ increases the region of cyclic dominance, but also eliminates the possibility of complete dominance of inspectors. Qualitatively, however, the evolutionary dynamics within the different phases does not change compared to $\gamma=0.5$, and remains the same also for $\gamma>1$. Dash-dotted gray lines in \textbf{A} and \textbf{B} correspond to the condition $p=\beta$, i.e. where the probability for criminals to be detected is the same as the temptation to commit crime, and a transition to criminal behavior would be expected according to the rational choice equation~(\ref{one}).}
\label{phase}
\end{figure}

For comparison, let us first summarize the behavior of the traditional two-strategy inspection game in a non-spatial setting. Assuming that individuals interact randomly and spread relative to their payoffs, the governing replicator equation \cite{hofbauer_98} predicts that the probability of committing a crime is $c/r=\alpha/\gamma$ and the probability of inspection is $g/f=\beta$. Accordingly, as soon as $\alpha \geq \gamma$ the whole population is dominated by criminals, while for $\alpha<\gamma$ criminals and inspectors coexist in different proportions depending on $\beta$. These basic considerations indicate a relatively simple and smooth dependence on the model parameters $\alpha$, $\beta$, and $\gamma$, but they also suggest that higher punishment implies less inspection, which sounds questionable and is in fact frequently subject to critique \cite{rauhut_jasss09}.

The collective behavior of the proposed three-strategy spatial inspection game is much more complex, and also counter-intuitive. In contrast to the traditional case of well-mixed interactions briefly discussed before, we find various phase transitions between different kinds of collective outcomes, as demonstrated in Fig.~\ref{phase}. We find four different phases:
\begin{enumerate}
\item a dominance of criminals for high temptation $\beta$ and high inspection costs $\alpha$,
\item a coexistence of criminals and punishing inspectors (the $P+C$ phase) for large values of temptation $\beta$ and moderate inspection costs $\alpha$,
\item a dominance of police for moderate inspection costs $\alpha$ and low values of temptation $\beta$, but only if the inspection incentives $\gamma$ are moderate [see panel (A)], and
\item cyclical dominance for small inspection costs $\alpha$ and small temptation $\beta$, where criminals outperform ordinary individuals, while these outperform punishing inspectors, and those win against the criminals (the $C+O+P$ phase).
\end{enumerate}
The phase transitions are either continuous (solid lines in Fig.~\ref{phase}) or discontinuous (dashed lines in Fig.~\ref{phase}), and this differs markedly from what would be expected according to the rational choice equation~(\ref{one}) or the well-mixed model due to the significant effects of inspection and spatial interactions.

Starting in the cyclical dominance $C+O+P$ phase and going clockwise around the phase diagram depicted in Fig.~\ref{phase}(A), we find that, as the temptation $\beta$ increases towards the lower solid line, the invasion of criminals into area of the ordinary individuals becomes more and more effective. This is expected because the profits of criminals increase with increasing values of $\beta$. Consequently, at the transition point, ordinary individuals die out. The ordeals of ordinary individuals, however, negatively affect also the criminals because there is increasing shortage of those left to invade. This is counter-intuitive, but in fact common in spatial models of cyclical interactions \cite{szolnoki_pre11}, where the fierce competition between two strategies (in our case ordinary individuals and criminals) frequently leads to the flourishing of the third strategy (in our case the inspectors). The overlayed color map conveys clearly that this effect indeed drastically lowers the crime rate near the $C+O+P \to P+C$ transition line. Once in the $P+C$ phase, however, increasing values of $\beta$ add strength to the criminals, which makes it increasingly difficult for the inspectors to contain crime. The crime rate therefore increases until eventually the pure $C$ phase is reached through the second continuous phase transition (upper solid line).

\begin{figure}
\centerline{\epsfig{file=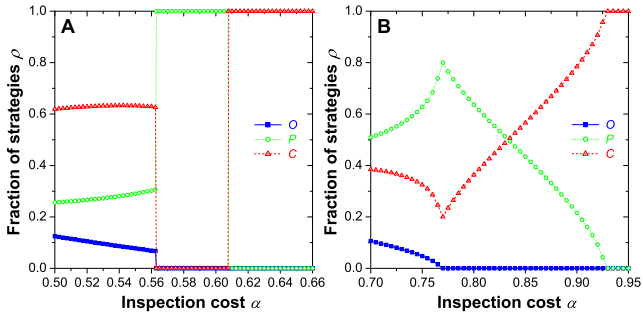,width=8.5cm}}
\caption{Representative cross-sections of phase diagrams, invalidating straightforward gain-loss principles (``linear thinking'') as proper description of the relationship between crime, inspection and punishment rates. As the inspection costs $\alpha$ increase, there are virtually no clear trends regarding the impact that such changes have on the outcome of the evolutionary process. (\textbf{A}) For $\beta=0.2$ and $\gamma=0.5$, increasing $\alpha$ initially leaves the stationary densities of strategies almost unaffected, while a discontinuous first-order phase transition to complete dominance of inspectors occurs at $\alpha=0.5625$. Another discontinuous first-order phase transitions follows at $\alpha=0.6075$, where the dominance of inspectors is replaced by the dominance of criminals. (\textbf{B}) For $\beta=0.4$ and $\gamma=1.0$, ordinary people first disappear in a second-order continuous phase transition at $\alpha=0.765$, thereby terminating the cyclic $C+O+P$ phase. In the following mixed $P+C$ phase, the impact of increasing inspection costs $\alpha$ leads to an increase in the number of criminals, which finally gives rise to a complete dominance through another second-order continuous phase transition occurring at $\alpha=0.925$. It is worth noting that the impact of increasing temptation $\beta$ at a fixed value of $\alpha$ is analogous.}
\label{cross}
\end{figure}

Continuing clockwise from the pure $C$ phase, we enter into the pure $P$ phase by means of a discontinuous first-order phase transition that emerges suddenly for sufficiently low values of the temptation $\beta$ and the inspection costs $\alpha$. The reason for the discontinuous first-order phase transition is an indirect competition between the inspectors and the criminals, the mediators of which are the ordinary individuals. If the inspectors succumb to ordinary individuals before forming a sufficiently large compact cluster, the criminals are subsequently able to eliminate all the ordinary individual to form the pure $C$ phase. If, on the other hand, inspectors are able to prevail long enough for criminals to eliminate the ordinary individuals, the final battle between the inspectors and the criminals is won by the former, thus yielding a pure $P$ phase. The transition between these two outcomes is sudden and imminent, and it can be triggered either by lowering the value of $\beta$ or the value of $\alpha$ (or both).

As the inspections costs $\alpha$ are lowered further, however, the pure $P$ phase gives way to the $C+O+P$ phase via another discontinuous first-order phase transition. The latter emerges because, at sufficiently low values of $\alpha$, the inspectors can survive alongside ordinary individuals long enough to establish a cyclic dominance relation together with the criminals. As described before, in the $C+O+P$ phase ordinary individuals invade the inspectors, inspectors invade the criminals, who in turn invade the ordinary individuals.

Notably, for $\gamma=1$ [see Fig.~\ref{phase}(B)] the phases and the evolutionary outcomes are very similar, with the only difference that the pure $P$ phase does not emerge. This is because higher values of $\gamma$ act similarly on the inspectors as lower values of $\alpha$. Accordingly, compared to the $\gamma=0.5$ [see Fig.~\ref{phase}(A)] case, the phase transition lines move towards larger values of $\alpha$, and the intricate indirect competition between the inspectors and the criminals always results in either a pure $C$ phase or a mixed $P+C$ phase, which is separated from the $C+O+P$ phase.

\begin{figure}
\centerline{\epsfig{file=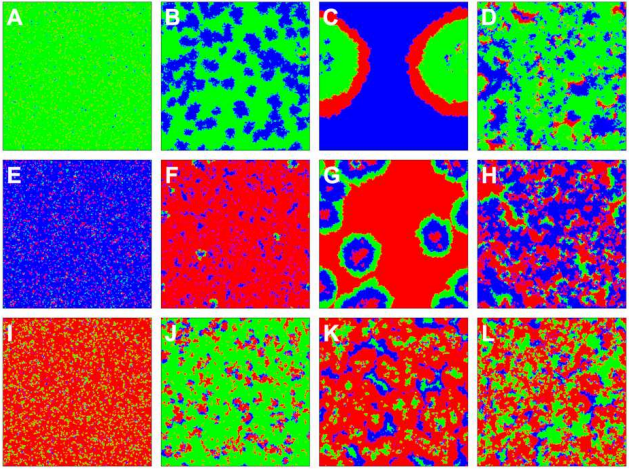,width=8.5cm}}
\caption{Snapshots of three different realizations of the cyclic $C+O+P$ phase, revealing the microscopic dynamics behind the cycles of crime. Invasion fronts can be either led by the criminals (red), by the inspectors (green), or by ordinary individuals (blue). (\textbf{A-D}) For small inspection costs $\alpha=0.05$, large temptation $\beta=1.0$ and medium inspection incentives $\gamma=0.5$, the situation is initially dominated by inspectors, after which clusters of ordinary individuals spread. Then, a front of criminals invades the area of ordinary individuals, but they are chased by inspectors. As a result, inspectors prevail, with clusters of ordinary individuals and a few criminals in between. (\textbf{E-H}) For medium inspection costs $\alpha=0.2$, small temptation $\beta=0.05$ and medium inspection incentives $\gamma=0.5$, the situation is first dominated by ordinary individuals, then by criminals. Afterwards, inspectors form an invasion front entering the domain of criminals, followed by ordinary individuals. Eventually, the situation is dominated by ordinary individuals with some inspectors and criminal enclaves in between. (\textbf{I-L}) For even higher inspection costs $\alpha=0.5$, moderate temptation $\beta=0.25$ and medium inspection incentives $\gamma=0.5$, the evolutionary process is dominated by criminals in the early stage and later by inspectors. Finally, however, inspectors and criminals prevail, with a few clusters of ordinary individuals in between. Supplementary videos showing all three evolutionary processes from the beginning until the convergence to the stationary distribution of strategies are available at the following links: (\textbf{A-D}) \href{http://www.youtube.com/watch?v=pH9l-2h6PRo}{\textcolor{blue}{youtube.com/watch?v=pH9l-2h6PRo}}, (\textbf{E-H}) \href{http://www.youtube.com/watch?v=gVnCN3a9ki8}{\textcolor{blue}{youtube.com/watch?v=gVnCN3a9ki8}} and (\textbf{I-L}) \href{http://www.youtube.com/watch?v=ehlDSde3BM4}{\textcolor{blue}{youtube.com/watch?v=ehlDSde3BM4}}.}
\label{snaps}
\end{figure}

Figure~\ref{cross} features a more detailed quantitative analysis, displaying in panel (A) the sudden first-order phase transitions from the $C+O+P$ phase to the pure $P$ phase, and from the pure $P$ phase to the pure $C$ phase as inspection costs $\alpha$ increase. In panel (B), we can observe the second-order phase transitions, first from the $C+O+P$ phase to the $P+C$ phase, and then from the $P+C$ phase to the pure $C$ phase. It follows that the exact impact of each specific parameter variation depends strongly on the location within the phase diagram, i.e. the exact parameter combination. General statements like ``increasing the fine reduces criminal activity'' tend to be wrong, which contradicts the established point of view. This may explain why empirical evidence is not in agreement with existing theoretical expectations, and it also suggests we ought to reconsider the common perspective on crime. Our model, although minimalist, allows to conclude that ``linear thinking'' brakes down as a means to devise successful crime prevention policies.

Snapshots presented in Fig.~\ref{snaps} further underline the complex evolutionary dynamics that underlies the prevention of crime. Depending on the parameter values, minute but compact clusters of the seemingly defeated strategy [criminals in panel (B), inspectors in panel (F), ordinary individuals in panel (J)] can resurrect, forming invasion fronts that are characteristic for cyclic dominance. Eventually, a dynamical equilibrium is reached in which all three strategies coexist in varying abundance.

It is also possible to derive implications for vanishing values of $\alpha$ and $\beta$ that are very difficult to capture by means of simulations, and which may constitute the starting point for a separate study. For example, in the limit of vanishing (relative) inspection costs $\alpha$ and vanishing (relative) temptation $\beta$ (i.e. $\alpha=\beta=0$), inspectors outperform criminals, and a mixed phase of ordinary individuals and inspectors remains. As both strategies receive equal payoffs for $\alpha=\beta=0$, the finally winning strategy is the outcome of a logarithmically slow coarsening process \cite{dornic_prl01}. The victor is usually the strategy that occupies the larger portion of the lattice at the time of extinction of the first strategy (the criminals). However, if the inspection costs $\alpha$ increase only slightly while the temptation $\beta$ stays zero, the equivalence of ordinary individuals and inspectors is immediately broken in the favor of the former, so that ordinary people tend to dominate quickly. Notably, for $\beta=0$, the strategies of ordinary individuals and criminals are also equivalent. Therefore, the winner of the struggle between ordinary individuals and criminals is again determined by logarithmically slow coarsening if, by any chance, inspectors die out before they outperform the criminals. Increasing the temptation $\beta$ slightly above zero does not affect the competition between ordinary individuals and inspectors at low inspection costs $\alpha$. Yet, it helps criminals to convert more and more ordinary individuals, before they are finally overwhelmed by the evolutionary pressure of the inspectors. Accordingly, for small values of temptation $\beta$, inspectors will most likely be the winners of the logarithmically slow coarsening.

\begin{figure}
\centerline{\epsfig{file=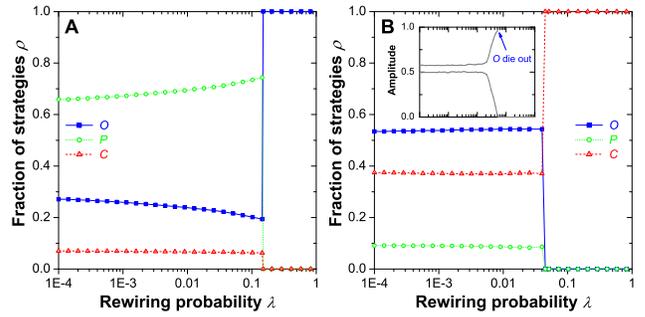,width=8.5cm}}
\caption{Robustness of crime cycles against the variation of network topology. The social interaction networks were constructed by rewiring links of a square lattice of size $400 \times 400$ with probability $\lambda$. For low values of $\lambda$, small-world properties emerge, while for $\lambda \rightarrow 1$ we have a random regular graph. As $\lambda$ is small and increases, the stationary fractions of the three competing strategies remain almost the same. However, due to the increasing interconnectedness of the players, the amplitude of oscillations increases. When a critical threshold value of $\lambda$ is reached, the maxima become comparable to the system size and oscillations terminate abruptly. The winner is the strategy that mediates the evolutionary competition between the two other strategies. (\textbf{A}) For small inspection costs $\alpha=0.05$, large temptation $\beta=1.0$ and moderate inspection incentives $\gamma=0.5$, ordinary people are the winners [see the evolution in panels (A-D) in Fig.~\ref{snaps}, in particular panel (C)]. (\textbf{B}) For moderate inspection costs $\alpha=0.2$, small temptation $\beta=0.05$, and medium inspection incentives $\gamma=0.5$, criminals are the winners [see the evolution in panels (E-H) in Fig.~\ref{snaps}, in particular panel (G)]. While cycles of crime are in general robust to variations of the network structure, the globalization by shortcut links adds another layer of complexity to the game that can result in the emergence of discontinuous phase transitions to absorbing states, for example, the prevalence of ordinary individuals (but not necessarily so). Note, however, that the evolutionary dynamics becomes more and more fragile as the cycles escalate [inset in (\textbf{B})] shows the envelope of oscillations of $\rho_O$), until they eventually involve almost the whole population. A supplementary video depicting such an evolutionary process where criminals are the victors is available at \href{http://www.youtube.com/watch?v=oGNOmLognOY}{\textcolor{blue}{youtube.com/watch?v=oGNOmLognOY}}. The final outcome of this dynamics may be hard to predict, especially if the population size is small and the strategies become subject to random extinction.}
\label{nets}
\end{figure}

Last but not least, we have also studied the relevance of the network structure. Figure~\ref{nets} present results for a periodic grid, in which a certain fraction $\lambda$ of links has been randomly rewired, thereby creating a regular small-world network for small values of $\lambda$ and a regular random network for $\lambda=1$ \cite{szabo_jpa04}. It turns out that the conclusions for these interaction networks stay qualitatively the same. However, between 5\% and 10\% of shortcut links, we find discontinuous first-order phase transitions, which are related to a sudden extinction of two strategies. Which two these are depends on the location within the $C+O+P$ phase, and is directly related to the highest peaks of abundance during the oscillatory dynamics. In fact, beyond a certain value of $\lambda$, shortcuts drastically increase the amplitude of oscillations, to the point that it becomes system-wide regardless of the system size, as depicted in the inset of Fig.~\ref{nets}(B). Extinction thus becomes inevitable, which gives rise to abrupt changes in the evolutionary dynamics, and it creates shocks of abundance of either criminals, inspectors or ordinary individuals. These results also indicate that the social structure of interaction networks adds another layer of complexity to crime containment. This does not have to do solely with interaction among us being limited rather than well-mixed, but also with the few long-range interaction creating globalization effects that emerge suddenly and without any (or very subtle) prior warning.

\section{Discussion}
Crime remains one of the plagues of society despite best efforts to fight it. According to classical rational choice theory, crime should disappear when the punishment, multiplied with the probability to get fined, exceeds the benefits of a crime. Hence, large enough fines should be able to eliminate crime. Nevertheless, crime occurs in all societies, independently of how harsh the punishment may be. A new approach may, therefore, be needed. Our paper discusses the possible emergent/system-immanent nature of crime. Based on a spatial evolutionary game theoretical model, it studies a situation in which individuals can engage in three kinds of activities: to commit a crime, to inspect and punish, or to do neither of both. The model considers the level of temptation to commit a crime as well as the inspection costs and the punishment intensity. We discover a non-monotonous dependence of the system behavior and level of crime as a function of the temptation and inspection costs as compared to the punishment intensity. In particular, we find sudden transitions to unexpected system behaviors at certain tipping points. This counter-intuitive result can only be understood as outcome of an emergent, collective dynamics, as it can occur in complex evolutionary systems.

Studying the evolutionary dynamics of crime makes some empirical features better comprehensible, which cannot be explained with conventional models of crime. This particularly concerns the circumstance that higher punishment fines do not necessarily imply less crime. In fact, the contrary could happen. Evolutionary game-theoretical models may thus provide unique insights into how the ``social immune system'' to fight crime actually works, which is a critical prerequisite for the development of new and improved crime containment strategies. For example, while the use of surveillance technologies may reduce relative inspection costs, our results suggest that they are unable to eliminate crime. Besides inspection costs, inspection incentives matter as well. A lot of crime and corruption happens, because incentives to reveal them are too low.

The results of our computer simulations suggest that criminals cannot be completely eliminated, when criminal activity creates a gain. On the other extreme, if the temptation $\beta$ is too high, the situation can get totally out of control, ending in a tragedy of the commons where everybody exploits everybody else. This situation has been described as primordial state of society with the words ``homo hominis lupus'' (``people are like wolves to each others'') \cite{hobbes_1651}. This state may indeed occur as result of the breakdown of public order, e.g. after large-scale disasters. To be controllable, the gain of criminal activity must stay below a certain critical threshold, and the relative inspection costs must be sufficiently low. Finding the right level of the punishment fine is crucial.

Furthermore, note that the inherent cyclical dynamics of crime, which becomes apparent in our study, might also lead to regulatory cycles \cite{skarin_sds09}. Cyclical changes between conservative and liberal times are well-known. In particular, they can be seen in the cycles of regulation and deregulation in the financial system. At first sight, the cyclical dynamics in our model appears to be qualitatively similar to the one governing social dilemma games involving cooperators, defector and loners \cite{hauert_s02, semmann_n03}, in particular the spontaneous emergence of cyclic dominance and discontinuous phase transitions \cite{szabo_prl02}. However, it must be emphasized that we get cyclical dynamics due to spatial interactions, not because of an in-built cyclical dominance between the three strategies. In fact, the role of ordinary individuals in our model is very different from that of loners. Loners are traditionally introduced as players abstaining from the game and being satisfied with a marginal return \cite{hauert_s07}. In contrast, ordinary individuals in our model are everything but that. They are the main protagonists (the ``masses''), catalyzing both, the rewards to criminals and the costs to inspectors. While a simplified two-strategy inspection game with criminals and inspectors provides some elementary insights, only the inclusion of a third strategy of ``ordinary people'' yields reasonably realistic results.

Although we recognize the minimalist nature of our model, it may eventually change our basic understanding of why, when and where crime is likely to occur, and how it may be more efficiently contained. In particular, it is important to recognize that crime is not simply the result of activities of criminals. It should be rather viewed as result of social interactions of people with different behaviors, the collective dynamics resulting from such interactions \cite{castellano_rmp09, bohorquez_n09, johnson_s11}, and the spatiotemporal social context they create. One might thus want to reconsider the common perspective on crime. The social environment seems to be quite important for the emergence/occurrence of crime, and models like ours can help to better explain why social context matters. In other words, crime may not be well understood by just assuming a ``criminal nature'' of particular individuals (the ``criminals'') -- this picture probably applies just to a fraction of people committing crimes. Our results suggest that changing the social context and conditions may be able to make a significant contribution to the reduction of crime. This would have relevant implications for policies and law.

Extensions of our model might be promising to answer questions such as the following: How can corruption or organized crime be modeled? What other strategies besides peer inspection and punishment are available, and how are they changing the evolutionary dynamics? For example, how does pool punishment (i.e. a public sanctioning system) change the game and its outcome? Would a reward system contain crime more effectively than a punishment system? And are reputation systems more efficient than surveillance approaches? These questions are particularly relevant at a time, where information and communication technologies make a Big Brother Society feasible and discussions whether more or less private weapons are creating a safer world are waiting for scientific answers.

\section{Methods}
We start Monte Carlo simulations of the evolutionary dynamics of the proposed spatial inspection game with uniformly distributed strategies, each with an initial fraction of 1/3. The stationary fractions of the three strategies on the square lattice are determined by means of a random sequential update comprising the following steps: First, a randomly selected player $x$ plays the inspection game with all its nearest neighbors, generating an overall payoff of $P_{s_x}$. Then, one of the nearest neighbors of player $x$ is chosen randomly, and this player $y$ is assumed to get the payoff $P_{s_y}$ analogously to the previous player $x$. Finally, player $y$ is assumed to imitate the strategy of player $x$ with probability
\begin{equation}
q=\frac{1}{1+\exp[(P_{s_y}-P_{s_x})/K]} \,  ,
\label{MNL}
\end{equation}
where $K$ determines the level of uncertainty in strategy adoptions and the inverse $K^{-1}$ represents the so-called intensity of selection. Equation (\ref{MNL}) corresponds to the empirically supported multinomial logit model \cite{mcfadden_74}, which for two decision alternatives is also known as Fermi law. We set $K=0.5$ to account for the traditional assumption that better performing players are imitated more frequently, although players might sometimes adopt a less successful strategy (e.g. due to uncertain information or trial-and-error behavior).

In each full time step of the game, all players adopt the strategy of one of their neighbors once on average. Depending on the proximity to phase transition points and the typical size of emerging spatial patterns, the linear system size $L$ is varied between 200 and 1600. Equilibration in a statistical sense takes up to $10^6$ full rounds of the game.

\begin{acknowledgments}
This research was supported by the Slovenian Research Agency (Grant No. J1-4055), by the Future and Emerging Technologies program FP7-COSI-ICT of the European Commission through the project QLectives (Grant No. 231200), and by the ERC Advanced Investigator Grant ``Momentum'' (Grant No. 324247).
\end{acknowledgments}


\begin{thebibliography}{10}

\bibitem{gurerk_s06}
Gurerk O, Irlenbusch B, Rockenbach B (2006) The competitive advantage of
  sanctioning institutions.
\newblock Science 312: 108--111.

\bibitem{sigmund_n10}
Sigmund K, De~Silva H, Traulsen A, Hauert C (2010) Social learning promotes
  institutions for governing the commons.
\newblock Nature 466: 861--863.

\bibitem{traulsen_prsb12}
Traulsen A, R{\"o}hl T, Milinski M (2012) An economic experiment reveals that
  humans prefer pool punishment to maintain the commons.
\newblock Proc R Soc London B 279: 3716--3721.

\bibitem{fehr_aer00}
Fehr E, {G\"a}chter S (2000) Cooperation and punishment in public goods
  experiments.
\newblock Am Econ Rev 90: 980--994.

\bibitem{gardner_a_an04}
Gardner A, West SA (2004) Cooperation and punishment, especially in humans.
\newblock Am Nat 164: 753--764.

\bibitem{rockenbach_n06}
Rockenbach B, Milinski M (2006) The efficient interaction of indirect
  reciprocity and costly punishment.
\newblock Nature 444: 718--723.

\bibitem{eldakar_pnas08}
Eldakar OT, Wilson DS (2008) Selfishness as second-order altruism.
\newblock Proc Natl Acad Sci USA 105: 6982--6986.

\bibitem{boyd_s10}
Boyd R, Gintis H, Bowles S (2010) Coordinated punishment of defectors sustains
  cooperation and can proliferate when rare.
\newblock Science 328: 617--620.

\bibitem{fehr_n02}
Fehr E, G{\"a}chter S (2002) Altruistic punishment in humans.
\newblock Nature 415: 137--140.

\bibitem{fowler_n05}
Fowler JH, Johnson T, Smirnov O (2005) Human behaviour: Egalitarian motive and
  altruistic punishment.
\newblock Nature 433: E1--E1.

\bibitem{fowler_pnas05}
Fowler JH (2005) Altruistic punishment and the origin of cooperation.
\newblock Proc Natl Acad Sci USA 102: 7047--7049.

\bibitem{lehmann_an07}
Lehmann L, Rousset F, Roze D, Keller L (2007) Strong reciprocity or strong
  ferocity? A population genetic wiev of the evolution of altruistic
  punishment.
\newblock Am Nat 170: 21--36.

\bibitem{fehr_n03}
Fehr E, Rockenbach B (2003) Detrimental effects of sanctions on human altruism.
\newblock Nature 422: 137--140.

\bibitem{boyd_pnas03}
Boyd R, Gintis H, Bowles S, Richerson PJ (2003) The evolution of altruistic
  punishment.
\newblock Proc Natl Acad Sci USA 100: 3531--3535.

\bibitem{henrich_s06b}
Henrich J, McElreath R, Barr A, Ensminger J, Barrett C, et~al. (2006) Costly
  punishment across human societies.
\newblock Science 312: 1767--1770.

\bibitem{milinski_n08}
Milinski M, Rockenbach B (2008) Punisher pays.
\newblock Nature 452: 297--298.

\bibitem{UCR}
\protect{Federal Bureau of Investigation's Uniform Crime Reporting Program}
  (2012) \href{http://www.ucrdatatool.gov/}{\textcolor{blue}{www.ucrdatatool.gov}}.

\bibitem{handbook}
Maguire M, Morgan R, Reiner R (2012) The Oxford Handbook of Criminology.
\newblock Oxford: Oxford University Press.

\bibitem{UNODC2}
\protect{UNODC (United Nations Office on Drugs and Crime)} (2011) Statistics on
  Crime, Robbery.

\bibitem{Raphael_2001}
Raphael S, Winter-Ebmer R (2001) Identifying the effect of unemployment on
  crime.
\newblock Journal of Law and Economics 44: 259--284.

\bibitem{Lin_2008}
Lin MJ (2008) Does unemployment increase crime? Evidence from U.S. data
  1974-2000.
\newblock Journal of Human Resources 43: 413--436.

\bibitem{LaFree_1999}
LaFree G (1999) Declining violent crime rates in the 1990s: Predicting crime
  booms and busts.
\newblock Annual Review of Sociology 25: 145--168.

\bibitem{Zimring_2007}
Zimring FE (2007) The Great American Crime Decline.
\newblock Oxford: Oxford University Press.

\bibitem{Curtis_1998}
Curtis R (1998) The important transformation of inner-city neighborhoods:
  Crime, violence, drugs and youth in the 1990s.
\newblock Journal of Criminal Law and Criminology 88: 1233--1276.

\bibitem{Johnson_2006}
Johnson BD, Golub A, Dunlap E (2006) The rise and decline of hard drugs, drug
  markets, and violence in inner-city New York.
\newblock In: Blumstein A, Wallman J, editors, The Crime Drop in America,
  Cambridge: Cambridge University Press. pp. 164--206.

\bibitem{LaFree_1998}
LaFree G (1998) Losing Legitimacy: Street Crime and the Decline of Social
  Institutions in America.
\newblock Boulder: Westview Press.

\bibitem{Barker_2012}
Barker V (2010) \protect{Explaining the Great American Crime Decline: A Review
  of Blumstein and Wallmann, Goldberger and Rosenfeld, and Zimring}.
\newblock Law \& Social Inquiry 35: 489--516.

\bibitem{Corsaro_2009}
Corsaro N, McGarrell EF (2009) \protect{Testing a promising homicide reduction
  strategy: re-assessing the impact of the Indianapolis ``pulling levers''
  intervention}.
\newblock J Exp Criminol 5: 63--82.

\bibitem{McGarrell_2010}
McGarrell EF, Corsaro N, \protect{Kroovand Hipple} N, Bynum TS (2010)
  \protect{Project Safe Neighborhoods and Violent Crime Trends in U.S. Cities:
  Assessing Violent Crime Impact}.
\newblock J Quant Criminol 26: 165--190.

\bibitem{Gomez_2006}
\protect{Gomez-Sorzano} GA (2006) Decomposing violence: Crime cycles in the
  twentieth century in the united states.
\newblock Applied Econometrics and International Development 7: 85--103.

\bibitem{Wilson_1995}
Wilson JQ (1995) Crime and public policy.
\newblock In: Wilson JQ, Petersilia J, editors, Crime, San Francisco: ICS
  Press. pp. 489--510.

\bibitem{Fox_1996}
Fox JA (1996) Trends in Juvenile Violence: A Report to the United States
  Attorney General on Current and Future Rates of Juvenile Offending.
\newblock Washington: Bureau of Justice Statistics.

\bibitem{doob_cj03}
Doob AN, Webster CM (2003) Sentence severity and crime: Accepting the null
  hypothesis.
\newblock Crime and Justice 30: 143--195.

\bibitem{becker_jpe68}
Becker GS (1968) Crime and punishment: An economic approach.
\newblock Journal of Political Economy 76: 169--217.

\bibitem{herrmann_s08}
Herrmann B, Thoni C, Gachter S (2008) Antisocial punishment across societies.
\newblock Science 319: 1362--1367.

\bibitem{dreber_n08}
Dreber A, Rand DG, Fudenberg D, Nowak MA (2008) Winners don't punish.
\newblock Nature 452: 348--351.

\bibitem{rand_s09}
Rand DG, Dreber A, Ellingsen T, Fudenberg D, Nowak MA (2009) Positive
  interactions promote public cooperation.
\newblock Science 325: 1272--1275.

\bibitem{rand_nc11}
Rand DG, Nowak MA (2011) The evolution of antisocial punishment in optional
  public goods games.
\newblock Nat Commun 2: 434.

\bibitem{hardin_g_s68}
Hardin G (1968) The tragedy of the commons.
\newblock Science 162: 1243--1248.

\bibitem{short_pre10}
Short MB, Brantingham PJ, \protect{D'Orsogna} MR (2010) Cooperation and
  punishment in an adversarial game: How defectors pave the way to a peaceful
  society.
\newblock Phys Rev E 82: 066114.

\bibitem{tsebelis_rs90}
Tsebelis G (1990) Penalty has no impact on crime: A game theoretic analysis.
\newblock Rationality and Society 2: 255--286.

\bibitem{inspect}
Avenhaus R, \protect{von Stengel} B, Zamir S (2002) Inspection games.
\newblock In: Aumann RJ, Hart S, editors, Handbook of Game Theory, Amsterdam:
  Elsevier. p.~51.

\bibitem{panchanathan_n04}
Panchanathan K, Boyd R (2004) Indirect reciprocity can stabilize cooperation
  without the second-order free rider problem.
\newblock Nature 432: 499--502.

\bibitem{fowler_n05b}
Fowler JH (2005) Second-order free-riding problem solved?
\newblock Nature 437: E8--E8.

\bibitem{helbing_ploscb10}
Helbing D, Szolnoki A, Perc M, Szab{\'o} G (2010) Evolutionary establishment of
  moral and double moral standards through spatial interactions.
\newblock PLoS Comput Biol 6: e1000758.

\bibitem{nowak_n92b}
Nowak MA, May RM (1992) Evolutionary games and spatial chaos.
\newblock Nature 359: 826--829.

\bibitem{nowak_pnas94}
Nowak MA, Bonhoeffer S, May RM (1994) Spatial games and the maintenance of
  cooperation.
\newblock Proc Natl Acad Sci USA 91: 4877--4881.

\bibitem{nowak_ijbc94}
Nowak MA, Bonhoeffer S, May RM (1994) More spatial games.
\newblock Int J Bifurcat Chaos 4: 33--56.

\bibitem{hauert_n04}
Hauert C, Doebeli M (2004) Spatial structure often inhibits the evolution of
  cooperation in the snowdrift game.
\newblock Nature 428: 643--646.

\bibitem{santos_pnas06}
Santos FC, Pacheco JM, Lenaerts T (2006) Evolutionary dynamics of social
  dilemmas in structured heterogeneous populations.
\newblock Proc Natl Acad Sci USA 103: 3490--3494.

\bibitem{ohtsuki_n06}
Ohtsuki H, Hauert C, Lieberman E, Nowak MA (2006) A simple rule for the
  evolution of cooperation on graphs and social networks.
\newblock Nature 441: 502--505.

\bibitem{szabo_pr07}
Szab{\'o} G, F{\'a}th G (2007) Evolutionary games on graphs.
\newblock Phys Rep 446: 97--216.

\bibitem{tanimoto_pre07}
Tanimoto J (2007) Dilemma solving by coevolution of networks and strategy in a $2 \times 2$ game.
\newblock Phys Rev E 76: 021126.

\bibitem{santos_n08}
Santos FC, Santos MD, Pacheco JM (2008) Social diversity promotes the emergence
  of cooperation in public goods games.
\newblock Nature 454: 213--216.

\bibitem{fu_pre09}
Fu F, Wang L, Nowak MA, Hauert C (2009) Evolutionary dynamics on graphs: Efficient method for weak selection.
\newblock Phys Rev E 79: 046707.

\bibitem{tarnita_jtb09}
Tarnita CE, Ohtsuki H, Antal T, Fu F, Nowak MA (2009) Strategy selection in structured populations.
\newblock J Theor Biol 259: 570--581.

\bibitem{tanimoto_pre12}
Tanimoto J, Brede M, Yamauchi A (2012) Network reciprocity by coexisting learning and teaching strategies.
\newblock Phys Rev E 85: 032101.

\bibitem{short_pnas10}
Short MB, Brantingham PJ, Bertozzi AL, Tita GE (2010) Dissipation and
  displacement of hotspots in reaction-diffusion models of crime.
\newblock Proc Natl Acad Sci USA 107: 3961--3965.

\bibitem{short_siam10}
Short MB, Bertozzi AL, Brantingham PJ (2010) Nonlinear patterns in urban crime
  - hotspots, bifurcations, and suppression.
\newblock SIAM J Appl Dyn Syst 9: 462--483.

\bibitem{brantingham_cr12}
Brantingham PJ, Tita GE, Short MB, Reid S (2012) The ecology of gang
  territorial boundaries.
\newblock Criminology 50: 851--885.

\bibitem{hofbauer_98}
Hofbauer J, Sigmund K (1998) Evolutionary Games and Population Dynamics.
\newblock Cambridge, UK: Cambridge Univ. Press.

\bibitem{rauhut_jasss09}
Rauhut H, Junker M (2009) Punishment deters crime because humans are bounded in
  their strategic decision-making.
\newblock Journal of Artificial Societies and Social Simulation 12: 1--23.

\bibitem{szolnoki_pre11}
Szolnoki A, Szab{\'o} G, Perc M (2011) Phase diagrams for the spatial public
  goods game with pool punishment.
\newblock Phys Rev E 83: 036101.

\bibitem{dornic_prl01}
Dornic I, Chat{\'e} H, Chave J, Hinrichsen H (2001) Critical coarsening without
  surface tension: The universality class of the voter model.
\newblock Phys Rev Lett 87: 045701.

\bibitem{szabo_jpa04}
Szab{\'o} G, Szolnoki A, Izs{\'a}k R (2004) Rock-scissors-paper game on regular
  small-world networks.
\newblock J Phys A: Math Gen 37: 2599--2609.

\bibitem{hobbes_1651}
Hobbes T (1651) Leviathan or the Matter, Form and Power of a Common Wealth
  Ecclesiastical and Civil.

\bibitem{skarin_sds09}
Skarin B, Pavlov O, Saeed K, Skorinko J (2009) Modeling the cycles of gang and
  criminal behavior.
\newblock In: Ford A, Ford DN, Anderson EG, editors, Conference Proceedings of
  the 27th International Conference of the System Dynamics Society. pp.
  1165--1184.

\bibitem{hauert_s02}
Hauert C, De~Monte S, Hofbauer J, Sigmund K (2002) Volunteering as \protect{Red
  Queen} mechanism for cooperation in public goods game.
\newblock Science 296: 1129--1132.

\bibitem{semmann_n03}
Semmann D, Krambeck HJ, Milinski M (2003) Volunteering leads to
  rock-paper-scissors dynamics in a public goods game.
\newblock Nature 425: 390--393.

\bibitem{szabo_prl02}
Szab{\'o} G, Hauert C (2002) Phase transitions and volunteering in spatial
  public goods games.
\newblock Phys Rev Lett 89: 118101.

\bibitem{hauert_s07}
Hauert C, Traulsen A, Brandt H, Nowak MA, Sigmund K (2007) Via freedom to
  coercion: The emergence of costly punishment.
\newblock Science 316: 1905--1907.

\bibitem{castellano_rmp09}
Castellano C, Fortunato S, Loreto V (2009) Statistical physics of social
  dynamics.
\newblock Rev Mod Phys 81: 591--646.

\bibitem{bohorquez_n09}
Bohorquez JC, Gourley S, Dixon AR, Spagat M, Johnson NF (2009) Common ecology
  quantifies human insurgency.
\newblock Nature 462: 911-914.

\bibitem{johnson_s11}
Johnson NF, Carran S, Botner J, Fontaine K, Laxague N, et~al. (2011) Pattern in
  escalations in insurgent and terrorist activity.
\newblock Science 333: 81--84.

\bibitem{mcfadden_74}
McFadden D (1974) Conditional logit analysis of qualitative choice behavior.
\newblock In: Zarembka P, editor, Frontiers in Econometrics, New York: Academic
  Press. pp. 105--142.

\end{thebibliography}
\end{document}